\documentclass[a4paper]{llncs}

\usepackage[T1]{fontenc}
\usepackage{amsmath}
\usepackage{ifpdf}
\usepackage[utf8]{inputenc}
\usepackage{xspace}
\usepackage{xcolor}
\usepackage[english]{babel}
\usepackage{pdfpages}
\usepackage[hyphens]{url}
\ifpdf
 \usepackage{hyperref}
\else
 \ifx\hrd\undefined
  \ifx\final\undefined\def\hrd{hypertex}\else\def\hrd{dvipdfm}\fi
 \fi
 \usepackage[\hrd]{hyperref}
 \if\hrd{hdvips}
 \usepackage[hyphenbreaks]{breakurl}
 \fi
\fi
\hypersetup{
  pdftitle={Licensing the Mizar Mathematical Library},
  pdfauthor={Jesse Alama, Adam Naumowicz, Michael Kohlhase, Lionel Mamane, Piotr Rudnicki, Josef Urban},
  pdfkeywords={free culture, open data, free licensing, formal mathematics, mizar},
  colorlinks=true,
  linkcolor=black,
  citecolor=black,
}

\DeclareMathOperator{\mand}{\land}
\def\systemname#1{\textsf{#1}\xspace}
\def\libname#1{\textsf{#1}\xspace}

\newcommand{\mizar}{\systemname{Mizar}}
\newcommand{\coq}{\systemname{Coq}}

\newcommand{\mml}{\libname{MML}}
\newcommand{\MML}{\libname{MML}}

\makeatletter

\newcommand{\Rmnum}[1]{\expandafter\@slowromancap\romannumeral #1@}
\makeatother

\title{Licensing the \mizar{} Mathematical Library\thanks{The final publication of this paper is
  available at www.springerlink.com}}
\titlerunning{Licensing the \MML} \author{Jesse Alama\inst{1} \and
  Michael Kohlhase\inst{2} \and Lionel Mamane\inst{3} \and Adam
  Naumowicz\inst{4} \and Piotr Rudnicki\inst{5} \and Josef
  Urban\inst{6}\thanks{The first author was funded by the FCT project
    ``Dialogical Foundations of Semantics'' (DiFoS) in the ESF
    EuroCoRes programme LogICCC (FCT LogICCC/0001/2007).  The sixth
    author was supported by the NWO project ``MathWiki a Web-based
    Collaborative Authoring Environment for Formal Proofs''.  The authors have worked together as members of the SUM Licensing Committee. Special thanks for advice to: Stephan Schulz, David Wheeler, Michael Spiegel, Petr
    Pudlak, Bob Boyer, Adam Pease, Timo Ewalds, Adam Grabowski,
    Czeslaw Bylinski, and Andrzej Trybulec.}}
\institute{Center for Artificial Intelligence;
  New University of Lisbon; \email{j.alama@fct.unl.pt} \and Computer
  Science, Jacobs University; \email{m.kohlhase@jacobs-university.de}
  \and \email{lionel@mamane.lu} \and Institute of Computer Science,
  University of Bialystok; \email{adamn@math.uwb.edu.pl} \and
  Department of Computing Science, University of Alberta;
  \email{piotr@cs.ualberta.ca} \and
  Institute for Computing and Information Sciences;  Radboud University Nijmegen\\
  \email{josef.urban@gmail.com} } \authorrunning{Alama et al.}
\tocauthor{Alama, Kohlhase, Mamane, Naumowicz, Rudnicki, Urban}

\begin{document}

\maketitle

\begin{abstract}
  The \mizar{} Mathematical Library (\MML) is a large corpus of
  formalised mathematical knowledge.  It has been constructed over the
  course of many years by a large number of authors and maintainers.
  Yet the legal status of these efforts of the \mizar{} community has
  never been clarified.  In 2010, after many years of loose
  deliberations, the community decided to investigate the issue of
  licensing the content of the \MML, thereby clarifying and
  crystallizing the status of the texts, the text's authors,
  and the library's
  long-term maintainers.  The community has settled on a copyright and license policy that
  suits the peculiar features of \mizar{} and its community.  In this
  paper we discuss the copyright and license solutions.
  We offer our experience in the hopes that the communities
  of other libraries of formalised mathematical knowledge might take
  up the legal and scientific problems that we addressed for \mizar.
\end{abstract}

\section{Introduction - Formal mathematics and its roots}
\label{sec:intro}


The dream of formal thinking and formal mathematics (and its giant
offspring: computer science) has a long and interesting history that
we can hardly go into in this paper. Briefly, formal mathematics
started to be produced in \emph{computer understandable} encoding in
late 1960s. The first significant text of formal mathematics was van
Benthem's encoding of Landau's \emph{Grundlagen der
  Analysis}~\cite{Jutting79} in AUTOMATH~\cite{Bruijn80}. Since then,
large bodies of formal mathematics have been created within the fields
of Interactive and Automated Theorem Proving (ITP, ATP).
(See~\cite{2006provers} for an extensive overview of the systems,
formal languages, and their libraries.)  

As these libraries grow and their contents get re-used in new,
possibly unimagined and unintended contexts, their legal status needs
to be clarified and formalised.  In this paper we discuss how this
problem was tackled in the case of the \mizar{} Mathematical Library.
In Section~\ref{What} we discuss formal
mathematical libraries in general and the target library of interest
for us, the \mizar{} Mathematical Library (MML \cite{mml}), and the problem of
classifying its content as code vs. text.  We discuss in
Section~\ref{related} how some basic licensing problems were tackled by
other formal mathematics projects.  
In
Section~\ref{issues-solutions} we survey the main issues we faced, and
our (sometimes incomplete) solutions to them.  We offer further
problems and future work in Section~\ref{sec:conclusion}. Our final
copyright/licensing recommendation is accessible
online\footnote{\url{https://github.com/JUrban/MMLLicense/raw/master/RECOMMENDATION}},
and the \mizar{} copyright assignment and licensing policy are now parts
of the \mizar{}
distribution.\footnote{\url{http://mizar.org/version/current/doc/Mizar_FLA.pdf}\\ \url{http://mizar.org/version/current/doc/FAQ}\\ \url{http://mizar.org/version/current/doc/COPYING.interpretation}\\ \url{http://mizar.org/version/current/doc/COPYING.GPL}\\ \url{http://mizar.org/version/current/doc/COPYING.CC-BY-SA}}

\section{What is a formal math library?}
\label{What}

A library of formal mathematics is a collection of ``articles'' that
contain formalised counterparts of everyday informal mathematics.  Our
interest here is on the \mizar{} Mathematical Library (MML).  Here we
discuss some of the historical background of this library and the
problems it poses for the license designer.

\subsection{Historical background of the MML}
\label{sec:historical-background}

The year 1989 marks the start of a systematic collection of
inter-referenced \mizar{} articles. 
The first three articles were included into a \mizar{} data base
on January~1 ---~this is the official date of starting the \mizar{}
Mathematical Library~---~\mml, although this name appeared later.  

The copyright for the \mml{} has been owned by the Association of
\mizar{} Users (SUM, in Polish Stowarzyszenie U\.{z}ytkownik\'{o}w Mizara)
anchored in Bia\l ystok. SUM is a registered Polish association whose
statute\footnote{\url{http://mizar.org/sum/statute.new.html}} states
that the SUM's aim is popularizing, propagating and promoting the
\mizar{} language. The copyright assignment has been required from the
authors (typically SUM members) by SUM when submitting articles to
MML. This was likely related to the early decision to build a large
re-usable library of formal mathematics, which would presumably be
re-factored many times by the core MML developers. Another reason
speaking for the copyright centralization was the fact that the
potential uses of such a library were initially also quite unclear:
note that MML was started so early that it precedes the World Wide
Web, Linux, Wikipedia, arXiv, and the massive development of free
software and free web resources like Wikipedia and arXiv in the last
two decades, and the related development of free licenses.

While the MML copyright owner was always clearly SUM, it was nowhere
stated what was covered by the copyright and there was no policy for
licensing the use of \mml{} content should someone request a
permission for use other than contributing to the library.  For
example, Josef Urban was not sure whether it was legal to include his
translation~\cite{DBLP:conf/mkm/Urban03} of parts of the \mml{} into
the TPTP library, which is used also for testing/benchmarking of
ATPs ---~a process with some potential for commercial use. As discussed
in the next section, the ATP translation also makes MML more
executable in certain sense, in a similar way as for example Prolog
program can be viewed as a database and as a program
simultaneously. Such potential uses add to the need for a clear
licensing policy.

\subsection{Two aspects of formal mathematics: code and text}
\label{sec:license-considerations}

Formal mathematical libraries present a number of problems for the
license designer.  One crucial question for deciding upon
a license for the \mizar{} Mathematical Library: Is a \mizar{}
article more like a journal article, or is it more like a piece of
computer code, or is it both? And could existing law be suboptimal in
treating code differently from mathematical texts?

Although we are focused on \mizar{}, we note in passing that other
interactive proof assistants, whose communities might want to take up
the problem of licensing their content, face this issue in different
ways.
Definitions and proofs in \coq{}, for instance,
have originally been rather
more like computer programs (at least, prima facie) than \mizar{}
texts\footnote{This is however also changing: Coq has become capable
  of handling mainstream (not just constructive) mathematics recently,
  has gained a declarative proof mode
  and one of the aspirations of the Math Components project is to make the
  Coq presentations accessible to mathematicians.}.

We also need to address the issue of what can be done with formal
mathematical texts.  There is considerable interest in extracting
algorithms from proofs of universal-existential theorems.  What is the
status of such extracted or derived products?  \mizar{} does not, on
its face, permit such a straightforward, immediate extraction of
algorithms from proofs. There are however several mechanisms which
bring \mizar{} very close to executable code:
\begin{enumerate}
\item Formal mathematical formulas (more precisely: clauses)
  immediately give rise to very real computation in the Prolog
  language. For example, the Prolog algorithm for reversing lists:
\begin{verbatim}
reverse_of([],[]).
reverse_of([H|T],Result):- 
   reverse_of(T,RevT),
   append(RevT,[H],Result).
\end{verbatim}
is just two mathematical clauses (formulas):
\begin{align*}
& rev\_of([],[]) \wedge\\
& \forall H,T,RevT,Result: rev\_of(T,RevT) \mand append(RevT,list(H,[]),Result) \\
& \rightarrow rev\_of(list(H,T),Result)
\end{align*}
The only difference between a Prolog program and a set of mathematical
clauses is that the order of clauses matters in the Prolog program.
\item In systems for automated theorem proving (ATPs), even this
  ordering difference typically no longer exists. Many ATPs would thus
  be really able to ``compute'' the reverse of a particular list, just
  with the two formulas above, given to them in arbitrary order. The
  MPTP system~\cite{Urban06} translates all \mizar{} formulas into a format
  usable by ATPs, and thus allows such computations to be made.
\item \mizar{} itself actually contains pieces of more procedural
  semantics, e.g.\ the ``registrations'' mechanisms (a kind of limited
  Prolog mechanism). These mechanisms add restricted Prolog-like
  directed search to otherwise less directed general \mizar{} proof
  search, in a similar spirit as the Prolog language adds a
  particular restrictions to the order in which (Horn) clauses are
  traversed, as opposed to ATPs that work with arbitrary clauses and
  regardless of ordering. Input to these mechanisms are again \mizar{}
  formulas in special (Horn-like) from.
\item In principle, one \emph{could} extract a mangled form of
  constructive content from the classical content of the \mizar{}
  Mathematical Library by applying, say, the G\"odel-Gentzen
  translation from classical to intuitionistic logic. After that, the
  Curry-Howard isomorphism between proofs and programs could again be
  used to give procedural meaning to \mizar{} proofs (not formulas as
  in the above cases).
\end{enumerate}
In short, mathematics lends itself to executable interpretation not
just via the Curry-Howard proofs-as-programs interpretation of
constructive mathematics (reasonably well-known in the formal
community),
and extensions of it to classical mathematics,
but also by implicit formulas-as-programs interpretations,
developed by the Prolog and ATP communities. It is a well known fact
that efficient Prolog computation is just a restriction of ATP proof
search, and ATP proof search can be used as (typically less efficient
than Prolog) computation too. These mechanisms are to a certain extent
present already in \mizar, and are fully available using a number of
systems via the MPTP translation.

The formal/informal distinction cannot be straightforwardly based the
ability for a machine to extract content/meaning.  For example,
Wikipedia is today used for data and algorithm extraction, used in
advanced algorithms by, for example, IBM Watson system~\cite{watson}.
With sufficiently advanced extraction algorithms (which we are clearly
approaching), many ``documents'' can become ``programs''.  (For an
interactive demonstration, see~\cite{watson-trivia-challenge}.)

Various other factors make contributions to the \mizar{} Mathematical
Library like computer code.  An entry in Wikipedia can stand alone as
a sensible contribution.  \mizar{} articles, however, do not stand
alone (in general), because it cannot be verified ---~or even
parsed~--- in the absence of the background provided by other articles.
With some background knowledge in mathematics, some
human-understandable meaning can be extracted from \mizar{} texts~\cite{COMSEQ_2.ABS}:
\begin{verbatim}
reserve n,m for Element of NAT;
reserve g for Element of COMPLEX;
reserve s for Complex_Sequence;
definition
 let s;
 attr s is convergent means
 ex g st for p be Real st 0 < p
  ex n st for m st n <= m holds |.s.m-g.| < p;
end;
\end{verbatim}
is evidently the conventional
$\exists\forall\exists\forall$-definition of a convergent sequence of
complex numbers.  However, the exact meaning of this text can be
specified only with reference to the environment in which this text is
evaluated.  The environment provides some type information, such as
that \texttt{0} is a real number, \texttt{<} is a relation among real
numbers, the curious-looking \texttt{|.s.m-g.|} is a real number (it
is the absolute value of the difference of the $n$th term $s_{n}$ and
$g$), etc.

Thus, libraries of formal mathematics are akin to libraries of
software.  Code that calls a library function cannot function
without the library.  Similarly, a formal mathematical article is not
``formal'' as it cannot be understood in the absence of the other formal
articles it imports.  The background formal library is used as a
declarative and procedural knowledge to derive (not just ``verify'')
the contents of a new formal mathematical article.

\subsection{Code and text licenses}
\label{sec:code-text}
Because of the dual nature of the formal mathematical texts ---~they are both
human-readable (particularly when written in \mizar) and machine-processable ---~it is possible that we are
dealing with a new kind of object.
The licensing situation
in the world of free works\footnote{``free'' refers not (only) to zero price (Latin gratis),
but to freedom (Latin liber;
however, the consecrated vernacular expression is ``libre'',
so we will use ``libre'').
That is, a work that anybody is free to use, share and improve.}
within the
two categories (executable software code on the one hand, and
documents for human consumption on the other hand) has clear
``winners''.  On the code side,
statistics~\cite{BDOSRCLD,FLOSSmolePublication,Wheeler_GPL} based on
web scraping of contents of large free/open source software
repositories (such as \href{http://www.sf.net/}{SourceForge} or
\href{http://freshmeat.net/}{freshmeat}) show that slightly more than
a half of FLOSS\footnote{FLOSS for \emph{free/libre/open-source software}.} code is under a variant of the
\href{http://www.gnu.org/}{GNU}
\href{http://www.gnu.org/licenses/gpl.html}{General Public License}:
Data from the FLOSSmole project~\cite{FLOSSmoleProject} as of March
2011 shows that out of a total of \(43\:470\) FLOSS projects tracked
by freshmeat, \(24\:366\) are licensed under a version of the GNU
GPL\footnote{This count includes various versions of the GNU GPL, as
  well as the combination of the GNU GPL with special linking
  exceptions; it does not include significantly more liberal licenses,
  such as the GNU Lesser/Library GPL.}.
Of the projects hosted on
SourceForge, \(110\:412\) use a variant of the GPL, out of
\(174\:227\) that use a license approved by the
\href{http://www.opensource.org/}{Open Source Initiative}.

On the document side, although we were not able to find comparative usage
statistics, in our experience Creative Commons accounts for most of the mind-share, although domain-specific licenses have fair
success within their domain.
See~\cite{freedomdefined-licenses,gnu-various-licenses}.

%


%

Be it only for this reason,
it seemed prudent to us to allow for an eventual future relicensing,
and for this to keep central copyright ownership of the MML.
But also, the practice of licensing of free/libre electronic documents
is rather younger, and less mature, than free software licensing, thereby
increasing the ``risk'' that relicensing may be necessary
in the future.
For example, Wikipedia migrated
from a GNU Free Documentation license
to a Creative Commons BY-SA license
as recently as in 2009,
because a majority of other wikis had, by and large,
settled for a Creative Commons license,
and Wikipedia wished to make interchange of content
between Wikipedia and other wikis legally possible (and easy).

An additional uncertainty arises from the fact that MML articles
can be seen both as documents and as executable code;
possibly difficulties could arise at some point from this dual nature.
For example,
someone wanting to make a use of the MML
that sits squarely neither on the one side
nor the other,
but makes use of the duality in some way.
Possibly this use would neither be clearly
allowed by a license meant for executable code,
nor clearly allowed by a license meant for documents.
It is thus prudent to have a central authority
that can authorise such uses on a case-by-case basis
as they arise,
or revise the license of the MML once the issues
are better understood.

\subsection{Patents}\label{sec:patents}
Restrictions arising from patents are potentially just as lethal as
copyright restrictions for keeping a work free to use and
enhance by anybody for any purpose. The expected content of the MML
is however more of an abstract nature than of a technical nature.
However, first, (theoretically) only technical ideas are subject to
patents, notwithstanding the situation concerning software patents.
Second, the kind of things that the MML is now typically used for
would not infringe on a patent, even if the idea expressed in an MML
article would be covered by a patent: a patent do not forbid the
activity proper of studying or enhancing upon the covered idea; for
example, the RSA or IDEA algorithms being patented does not forbid
proving (formally or in paper mathematics) properties of these
algorithms, nor does it forbid teaching the algorithm publicly.  On
the contrary, the usual justification for the modern patent system is
to encourage inventors to make descriptions of their invention public,
rather than keeping them as trade secrets, so that such activities can
take place, and eventually lead to a larger and better exploitation of
the idea by society as a whole.  The activity forbidden by a patent is
manufacturing, selling or using an \emph{implementation} of the idea,
a machine based on the idea.  Eventual concrete advances in making a
specification of an algorithm in \mizar{} executable could create
interesting legal questions, but we are not yet at this
point\footnote{A similar question of where the boundary between
  ``implementation'' and ``human speech'' lies arose in the last years
  of the second millennium, concerning the CSS algorithm used to
  scramble the contents of Video DVDs, although in this case the
  problem did not originate from a patent, but from a then-new
  copyright meta-protection law, the USA Digital Millennium Copyright
  Act.  See \url{http://www.cs.cmu.edu/~dst/DeCSS/Gallery/} for
  examples.} although some preliminary experiments\footnote{Michal Michaels
  (\url{http://mmitech.net/michael/cv}) added computational contents
  to \mizar{} schemes and was extracting Lisp code for binary
  arithmetics in his MSc thesis, University of Alberta 1996.
} were quite encouraging. 
Third,
in the context of free software
we take our inspiration from,
how to handle patents is ---~by far~--- not as consensual
as copyright licensing.
Free documents, our other inspiration, usually don't have to deal
with patent issues.
For the combination of these three reasons,
we did
not address any patent issue in our initial licensing recommendation to SUM.

However it is worth noting that in GPL version 3 
the FSF has started to address the issue of
software patents more concretely,
inserting following into the GPL v3 preamble:
\begin{quote}
Finally, every program is threatened constantly by software patents.
States should not allow patents to restrict development and use of
software on general-purpose computers, but in those that do, we wish to
avoid the special danger that patents applied to a free program could
make it effectively proprietary.  To prevent this, the GPL assures that
patents cannot be used to render the program non-free.
\end{quote}
Section 11 of the GPL v3 text itself contains a blanket patent license
from every contributor in addition to the usual copyright license.  In
other words, GPL version 3 has the share-alike (transitive) aspect not
only with respect to copyright, but also with respect to patents.
However our copyright assignment setup means that a contributor not
redistributing his MML modifications himself never agrees to the GPL,
and thus evades its patent license provisions.  Additionally as the
Creative Commons license does not address patents, our dual-license
model allows redistributors of modified versions to avoid giving a
public patent license by choosing CC-BY-SA.  Addressing patent issues
(if relevant at all) is thus possible future work for us, probably
subject to some discussions with the authors of the FSFE Fiduciary
License Agreement (copyright assignment contract).

\section{Related licensing models}
\label{related}

For detailed overview of formal systems' licensing, see David
Wheeler's enumeration~\cite{Wheeler_Formal}. The tendency in academic
institutions over the last decade seems to go from
closed/non-free/non-commercial/unclear licensing terms on formal
systems, towards open/free/clear ones. Two examples are the SPASS
theorem prover from MPI Saarbr\"ucken, and the PVS verification system
from SRI. SPASS went from a custom license allowing only
non-commercial use (SPASS 1.0) to GPL2 (SPASS 2.0) to FreeBSD license
(SPASS 3.5). PVS has switched from a former commercial license to GPL
as of December
2006.\footnote{\url{http://pvs.csl.sri.com/mail-archive/pvs-announce/msg00007.html}}
One of the early rules (since 1997) of the CADE ATP System Competition
(CASC) has been that ``Winners are expected to provide public access
to their system's source
code'',\footnote{\url{http://www.cs.miami.edu/~tptp/CASC/14/Call.html\#Conditions}}
and that the systems' sources are after the competition regularly published
by the CASC organisers. Note that the situation is quite
different in the world of more applied formal tools like SAT and SMT
solvers. For example, neither the Z3 (Microsoft Research) nor the
Yices (SRI) SMT solvers are FLOSS.

However, our particular interest are not the formal systems per se,
but rather the formal libraries associated with them. As discussed in
Section~\ref{What}, this distinction ---~between the systems and
the mathematics formalised inside them ---~might be (im)possible to
various extents. For example, the HOL (Light) formalizations (and thus
the large mathematical Flyspeck project formalizing the proof of the
Kepler conjecture) are written directly in the ML (OCaml) programming
language. This is probably best captured as ``proof programming'' (the
ML code) over ``mathematical terms'' (specially parsed parts of the ML
code). Obviously, arbitrary programs (ML functions) thus are part of
the ``procedural proofs'' written in HOL (Light).\footnote{One might
  of course argue that the arbitrary ML functions in HOL (Light) serve
  not as ``parts of proofs'' but rather as ``proof generators'', i.e.,
  that the ``real proof'' is just the low-level HOL proof object
  (checked by HOL's LCF-like microkernel), which the user typically
  never sees. This is however a bit like saying that the Lisp (or C)
  macro language is not really a part of Lisp (or C) programming, or
  even that Lisp (or C) is just a ``program generator'', and the
  ``real program'' is just the compiled machine code.} On the other
hand, in \mizar, the distinction between the system's code (written in
Pascal) and the formalization code (written in \mizar) is very clear:
no Pascal programming is allowed inside the declarative mathematical
proofs.

\subsection{Licensing Models of Formal Libraries}

\begin{table*}[tbp]
  \caption{Overview of licenses of selected (formal) mathematical libraries}
  \begin{tabular}{|l||l|}
    \hline
    Library (System)&License\\
    \hline \hline
    Coq Standard Library (Coq) & LGPL \\
    \hline    
    Coq Repository at Nijmegen (Coq) & GPL \\
    \hline
    Math Components (Coq+ssreflect) & Not publicly available \\
    \hline
    Archive of Formal Proofs (Isabelle) & BSD or LGPL \\
\hline
    Isabelle Standard Libraries (Isabelle) & BSD \\
    \hline    
    HOL Light Standard Library (HOL Light) & BSD/MIT-like license \\
    \hline
    Flyspeck (HOL Light) & MIT license? \\
    \hline
    Wikipedia, PlanetMath & CC-BY-SA\\
    \hline
    SUMO & GPL + additional restriction\\
    \hline
    arXiv & CC-BY(-NC-SA) or public domain\\
          & or only arXiv allowed to distribute\\
    \hline
  \end{tabular}
\label{Licenses}
\end{table*}

All major formal libraries have so far used code licenses,
while informal ones like arXiv, PlanetMath and Wikipedia use document
licenses.  The GNU Free Documentation License (GFDL) has been
previously used by Wikipedia and PlanetMath, however, as Stephan
Schulz (a Wikipedia administrator) noted:
\begin{quote}
  The GFDL certainly is a reasonable choice. However, it has some
  warts, and large collaborative projects (in particular Wikipedia)
  have been moving (with support from the Free Software Foundation) to
  the Creative Commons CC-BY-SA license.  The CC-BY-SA license allows
  redistribution and changes, but requires maintaining the license and
  recognizing the contributors.
\end{quote}
Table~\ref{Licenses} summarises the licenses of several well-known
formal mathematical libraries, together with some major informal ones
like PlanetMath, Wikipedia, and arXiv.

Also note that the licensing differences between the formal libraries
can already now cause nontrivial problems. For example, the Coq
Repository at Nijmegen (CoRN) is an advanced mathematical library,
which however contains also a number of items that can be generally
useful to any Coq formalization. Thus, a credible scenario is that
pieces of CoRN might be gradually moved to the Coq Standard Library
(distributed with the Coq system). That however is not automatically
possible, since CoRN is licensed under GPL, which is stronger (more
restrictive) than LGPL (used by the Coq Standard Library). Similar
situation might arise when moving the LGPL-licensed entries in the
Isabelle Archive of Formal Proofs (AFP) to the Isabelle standard
libraries (BSD). So while our initial idea was to possibly optimise
the MML license(s) also with respect to possible future transfers and
translations between various formal and informal libraries, a survey
of the current situation revealed that this is hardly possible in the
existing chaos. This again leads us to the necessity of copyright
centralization, in order to be able to adapt to the likely future
changes of this chaotic global state of affairs.

\section{Issues and their solutions}
\label{issues-solutions}

In this section we discuss some of the problems we faced when
designing a license and copyright mechanism for the MML and how we addressed them.

\subsection{Features required from the license}

For the number of reasons mentioned above we wanted our solution for the MML copyright/license to give the SUM some control over
the current and future licensing, while at the same time not hindering legitimate ``open
science'' use of the MML, such as:
\begin{itemize}
\item translating its contents so that another proof assistant can use
  them;
\item archiving the MML to record the state of human knowledge;
\item allowing MML to be used for data-mining, mathematical search engines, and general AI systems;
\item benchmarking ATPs;
\item writing formal mathematics and publishing articles about it.
\end{itemize}

At the same time, we did want the SUM to have the authority to object
to uses of the MML that do not adhere to the rules of ``open science''
and block the free flow of ideas and results contained in the library.

Our solution was to adopt a fairly restrictive open license, adhering to
strong copyleft principles, with SUM as a rather powerful central
copyright owner.  We lay out a policy of ``ask and you shall be
allowed'' when it comes to certain uses of the MML that do not adhere
to our restrictive license,
but are within the rules of ``open science''.

\subsection{Linking and adaptation}

One consequence of viewing the MML as a collection of executable code
concerns the sensitive issue of \emph{linking}.  The MML is composed
of a large number of items that one can refer to, not unlike one using
a subroutine defined in some external library.  Thus, if someone has
proved the Jordan curve theorem in some form, one can use this
theorem
to prove some consequence, such as the
four-color theorem.  Likewise, one can use earlier definitions (such
as the definition of the power set operation or topological spaces) in
one's own work.  Such usage is analogous to linking by virtue of the
fact that one's text is not functional (or even meaningful) in the
absence of these earlier definitions,
and constitutes a derivative work of the used articles,
thereby triggering the share-alike mechanism of the GPL.

On the document side, the CC-BY-SA version 3.0 license~\cite{ccbysa}
share-alike mechanism is triggered
by the analogous notion of \emph{adaptation}.

A contribution to the MML naturally triggers the share-alike mechanism
of both licenses.
However, to dispel any doubt about this,
the MML licenses come with a binding interpretation note\footnote{\url{http://mizar.org/version/current/doc/COPYING.interpretation}}
that states that fact explicitly.

\subsection{Why open-source copyleft license?}

We settled upon a dual-licensing scheme based on the GPL version 3 \cite{gpl}
and CC-BY-SA version 3.0 \cite{ccbysa}.

The decision to adopt such a scheme was made with some reservation;
a dual-licensing scheme is evidently more complicated than a single
license.  However, the dual-license aspect does suit our situation
nicely owing to the dual nature of the MML as code and text.  The
intention is that the GNU GPL, which aims to cover computer code,
suits this aspect of the MML, whereas CC-BY-SA, which is designed to
cover texts (among other things) seems more appropriate when the MML
is considered as a collection of texts. The dual-license scheme is
good for \emph{adapting and copying} parts of MML. For example, it
allows to copy a piece of an MML proof into Wikipedia or PlanetMath,
which are licensed under CC-BY-SA. In such a case, the person copying
automatically chooses to use MML under the CC-BY-SA license. In the
same spirit, extraction of Prolog (or other) programs from MML
mentioned in Section~\ref{sec:license-considerations} would be covered
by GPL. The disadvantage of the MML dual-licensing is that
\emph{contributing to} MML gets difficult: contributors have to agree
both to GPL and CC-BY-SA. This was a difficult decision, however, once
we got that far, it allowed us to think even further and come up
(after discussions with the Software Freedom Law
Center\footnote{\url{http://www.softwarefreedom.org/}}) with copyright
assignment (see Section~\ref{why-keep}) as the best of the bad
solutions.


Our licenses feature \emph{strong copyleft protection}:
anything derived from one's contribution must be similarly freely
redistributable and enhanceable.
We believe that such transitivity promotes public
contribution, because a contributor can engineer his work safe in the
knowledge that his efforts,
and future enhancements to it,
cannot be taken away from him (or from
society),
and cannot be exploited for private gain
without contributing back to the common pot.

\subsection{Why keep the copyright ownership with SUM?}
\label{why-keep}


In our licensing model, we require that contributor to the MML assign
the copyright to their work to the SUM.

The main risk of mandatory copyright assignment is discouraging
potential contributors.  But since the MML has been functioning with
copyright assignment since its beginnings, the risk is mitigated.  The
community has already adopted this model, and probably will continue to
accept it.  At worst, mandatory copyright assignment might stunt the
future growth of the community.

To assuage this fear, following the models described in~\cite{kuhn_ca,hillesley_ca,rms_ca}, we recommended that:\footnote{Note that the last three points are now part of the Mizar Copyright Assignment.}
\begin{itemize}
\item  SUM be a relatively transparent association,
  and that it be open to contributors.
  One way
  in which SUM is open is through translation of its statutes into the
  major languages used for science and technology (currently, English).
  We also insisted that decisions taken by SUM be open to international
  members (i.e., not requiring physical Polish presence).
\item SUM pledges to maintain free (as in freedom) licensing
  of the assigned work.
\item SUM pledges that any profit made by the SUM from the work
  be used only for the advancement of science.
\item The copyright grant to the SUM is automatically rescinded
  if the SUM breaks the above pledges.
\end{itemize}

\subsection{Enforcing FLOSS}

Free/libre licenses had, and to a degree still have,
a reputation for being difficult, if possible at all,
to enforce, or the expose the licensor to abuse from the licensee.
This reputation, in our opinion, comes more from the fact
that enforcement (mainly by the FSF)
used to happen behind closed doors
rather than in a public forum,
and the final settlement typically would include
a ``no shaming'' clause
that kept the polite fiction
that the violator voluntarily complied with the GPL,
and never imagined doing otherwise,
much less did otherwise.
Eben Moglen, the general legal
counsel for the FSF, and Richard Stallman, the leader of the FSF,
used to say (publicly, during conferences) something to the effect of:
\begin{quote}
  The reason the GNU GPL has not been ``tested in court''
  is that each time the FSF threatens to sue over a GPL violation,
  the offender chooses to comply with the GPL
  rather than go to court.
  This, in essence, means that their legal counsel estimates
  their losing in court too probable to risk.
\end{quote}
More recently, some of the enforcement has become far more public, and
sometimes the public shaming is the main force behind the effort.  The
pioneer of this change is \url{gpl-violations.org}, created in 2004 by
Harald Welte to give GPL enforcement a faster and more dynamic pace
than the FSF's usual way of proceeding \cite{welte_fosdem2005}: the
FSF usually let violators continue their infringements for an interim
period of time, while the process of bringing them into compliance was
ongoing.

More strongly,
far from making the contents of the \MML{} more vulnerable to theft,
or limiting the freedom of the \mizar{} community, adopting a FLOSS
license such as the one we settled upon gives the community greater
strength.  There are cases where having a FLOSS license made a crucial
difference.  One of the earlier examples is the one of \texttt{g++},
the \systemname{C++} compiler in \systemname{GCC}, the GNU Compiler
Collection~\cite{FSFS}:
\begin{quotation}
Consider GNU C++. Why do we have a free C++ compiler? Only because the GNU GPL said it had to be free. GNU C++ was developed by an industry consortium, MCC, starting from the GNU C compiler. MCC normally makes its work as proprietary as can be. But they made the C++ front end free software, because the GNU GPL said that was the only way they could release it. The C++ front end included many new files, but since they were meant to be linked with GCC, the GPL did apply to them. The benefit to our community is evident.

Consider GNU Objective C. NeXT initially wanted to make this front end proprietary; they proposed to release it as .o files, and let users link them with the rest of GCC, thinking this might be a way around the GPL's requirements. But our lawyer said that this would not evade the requirements, that it was not allowed. And so they made the Objective C front end free software.
\end{quotation}
It is not inconceivable, as formal methods become more widely used,
that analogous cases could arise concerning the use of formalised
mathematical knowledge.

\subsection{What are reasonable conditions for copyright ownership?}

We settled for a copyright ownership agreement modeled on the FSFE
fiduciary license agreement (FLA)~\cite{fsfe}.  The FLA ``allows one
entity to safeguard all of the code created for a project by
consolidating copyright (or exclusive exploitation rights) to
counteract copyright fragmentation.''  In our case, the entity is SUM.
We opted for this kind of agreement to permit possible future
changes of the open-source licenses\footnote{The Wikipedia relicensing trouble being a strong motivation.} and also selected commercial activities benefiting scientific progress.  There already are projects (e.g., the NICTA
L4 project\footnote{\url{http://ertos.nicta.com.au/research/l4/}}) that are
sensitive to future commercial uses, while clearly benefiting the development of formal methods.
We chose the FSFE's FLA rather than the similar paperwork
used by e.g.\ the FSF mainly because
the FSFE's FLA is specifically written for European jurisdictions.

The main substantive change we made to the FLA is
allowing the SUM to sell commercial licenses to the MML,
when doing so is beneficial to progress in
science and technology,
subject to the restriction that proceedings must be used
to advance the SUM's goals,
namely popularising, propagating and promoting the \mizar{} language.
We see this way of proceeding as a tax levied
on people that want to benefit from science's production and advancement,
without playing by the rules of science of free interchange of ideas
and results.
In other words, people that want to use the results of science
without contributing their further enhancements back to the common pot.
This is somewhat similar to the ``polluter pays'' principle
that is becoming widespread in European and international
environmental law.

The fact that we set things up legally so that the SUM is
allowed to do so does not mean it has to;
if 
a majority of members (contributors) opposes it,
it won't happen and the MML will be, by and large,
\emph{unavailable} to people not willing
to contribute their enhancements to the common pot.

\section{Conclusion and Future Work}
\label{sec:conclusion}

In order to produce a suitable copyright and licensing model for the
MML, we delved into the question of what formal mathematics truly is.
We did not settle on a definitive answer yet, and it may well be
that (as often with legal concepts) the existing legal concepts
and preconceptions need to be updated as scientific progress is being
made.  The difference between executable code and (formal) mathematics
seems to be extremely tenuous, if it exists at all.
It is safe to say that
formal mathematical texts straddle a boundary between human
readability/consumability and machine readability/consumability.  An
even deeper question: what is formal knowledge, and what
is informal knowledge? 

Even when we handle this dilemma by dual licensing, there is no clear
winner with respect to the goal of making the MML compatible with as many
formal and informal mathematical libraries as possible.
The situation in this field
seems quite chaotic, and we hope that this paper will be of some help
to the developers of other formal libraries. In particular, our
recommendation in this chaotic situation is to centralise the
copyright ownership with trusted user associations, so that the
situation can be gradually improved by these bodies.

Although the MML has been licensed in a free way, the programs
that operate on these texts remain closed-source.  Ideally, both the
MML and these programs would be free/libre.  The license problem
here is simpler than the problem of licensing the MML, since we are
dealing simply with programs.
The process may end up
being gradual, starting with the \mizar{} parser. It is essential that such  a problem be tackled; the lack of any kind of open license for the whole of \mizar{} (its programs and its library), from a political standpoint of scientific research
being done for the general good of humanity
and available to all,
can push away some potential contributors toward other proof assistants and other libraries.\footnote{For example, the third author chose
\coq{} over \mizar{} for formalisation of surreal (Conway) numbers~\cite{concoq_types2004}
over such an issue,
although \mizar's set theory base
could have provided a more natural framework
for surreal numbers than \coq's type theory.}

For reasons discussed in Section~\ref{sec:patents}, we avoided the
problem of patents.  This is largely because we are not aware now of
how patents could play a role in formal mathematics.  In the future,
though, we may find that the subject needs to be revisited.

\nocite{mml}

\bibliographystyle{splncs03}
\bibliography{license}



\end{document}